\documentclass[12pt]{article}
\usepackage{latexsym,epsfig,epic}
\usepackage{rotating,wrapfig} 
\begin{document}
\setlength{\unitlength}{1mm}

\newcommand{\lp}{\left}
\newcommand{\rp}{\right}
\newcommand{\be}{\begin{equation}}
\newcommand{\ee}{\end{equation}}
\newcommand{\<}{\langle}
\renewcommand{\>}{\rangle}
\newcommand{\reff}[1]{(\ref{#1})}
\newcommand{\pslash}{\hbox{$p$\kern-0.5em\raise-0.35ex\hbox{$/$}}\thinspace}
\newcommand\BB{{\cal B}}
\newcommand\FF{{\cal F}}

\def\sfrac#1#2{{\textstyle\frac{#1}{#2}}} 

\title{Two-loop critical mass for Wilson fermions}

\author{
  { Sergio Caracciolo }              \\
  {\small\it Scuola Normale Superiore and INFN -- Sezione di Pisa}  \\[-0.2cm]
  {\small\it I-56100 Pisa, ITALIA}          \\[-0.2cm]
  {\small Internet: {\tt sergio.caracciolo@sns.it}}
  \\[-0.2cm]
  \and
  { Andrea Pelissetto}              \\
  {\small\it Dipartimento di Fisica and INFN -- Sezione di Roma I}
  \\[-0.2cm]
  {\small\it Universit\`a degli Studi di Roma ``La Sapienza"}        \\[-0.2cm]
  {\small\it I-00185 Roma, ITALIA}          \\[-0.2cm]
  {\small Internet: {\tt Andrea.Pelissetto@roma1.infn.it}}   \\[-0.2cm]
   \and
  { Antonio Rago }             \\
  {\small\it Dipartimento di Fisica and INFN -- Sezione di Torino} \\[-0.2cm]
  {\small\it Universit\`a degli Studi di Torino}        \\[-0.2cm]
  {\small\it Via Pietro Giuria 1, I-10100 Torino, ITALIA}          \\[-0.2cm]
  {\small Internet: {\tt Antonio.Rago@to.infn.it }}
   }

\maketitle

\thispagestyle{empty}   

\vspace{-1cm}
\begin{abstract}
We have redone a recent two-loop computation  of the
critical mass for 
Wilson fermions in lattice QCD by evaluating Feynman integrals 
with the coordinate-space method.  
We present the results for different types of infrared 
regularization.  We confirm both the
previous numerical estimates and the power of the coordinate-space
method whenever high accuracy is needed.
\end{abstract}

\clearpage

\section{Introduction}

The lattice formulation is at present the best tool for a 
nonperturbative study of QCD.
Its most widely used implementation is
due to Wilson~\cite{Wilson}. Here, the lattice fermion doubling 
problem 
is solved by the introduction of a formally irrelevant
second-order term  that gives a mass of the order of the cut-off
to the doublers. This term explicitly breaks chiral invariance and
therefore the limit of zero mass can be obtained only by a fine
tuning, which is to say that the bare and the renormalized quark 
mass no longer vanish together.
Nowadays, there are new formulations in which a form of
chiral invariance is preserved at the cost of a nonlocal
action (see, for example, the recent
reviews~\cite{Luscher,Neuberger,Creutz} and references therein). 
Even though this represents a very important progress
from a conceptual point of view, it is not clear if the new
formulations can become practical for numerical simulations.

In the Wilson formulation, the breaking of chiral invariance gives rise
to severe technical difficulties. Indeed, operators that have the correct 
chiral properties are obtained as sums of many different lattice
bare operators multiplied by suitable renormalization
constants. In principle, such constants should be computed 
nonperturbatively. In practice, this is often difficult
and thus one resorts to perturbation theory. 

Lattice perturbation theory is particularly complicated because
of the loss of Lorentz invariance and at present a two-loop
calculation represents a formidable task. For one-loop
computations
a completely general algebraic algorithm was introduced in 
Ref.~\cite{Menotti} for the pure gauge theory and extended 
in Ref. \cite{Burgio,Burgio:lat} to include Wilson fermions. 
This algorithm allows to express every one-loop
integral with gluon and Wilson-fermion propagators in terms of a
small number of basic constants that can be computed with
arbitrary precision~\cite{Burgio}. 
For two-loop calculations a powerful method was 
introduced by L\"{u}scher and Weisz~\cite{LW}
for the pure gauge theory, the so-called 
{\em coordinate-space} method. Its generalization 
to full QCD is again a nontrivial task. 
Some general ideas were presented in Ref.~\cite{Capitani} 
and will be extensively illustrated
elsewhere \cite{Capitani-prep}. 

Here, we present the first two-loop computation 
with the coordinate-space method in lattice QCD 
with fermions. The purpose is twofold. First,
we wish to check that the 
additional complications that are present in the method 
in the presence of fermions do not spoil 
its performance. Second, we want to compare it with 
the alternative momentum-space method, in which 
one works directly in momentum space. For this 
purpose, we repeat the calculation
of the critical mass for Wilson
fermions reported in Ref.~\cite{Follana}. 
It turns out that the coordinate-space method 
is quite precise, providing numerical expressions for the 
Feynman integrals with a precision of $10^{-10}$ or higher.
The momentum-space method also works well: 
although not as precise as the coordinate-space method, 
still the final results have a relative precision of 
$10^{-5}$. 

In this paper, we will consider Wilson fermions.
The dressed inverse fermion propagator has the form
\be
S^{-1}(p,m_B) =  i\,\overline{\pslash}\, +  m_B + M_W(p)
- \Sigma^L(p,m_B,g_0), 
\ee 
where, setting the lattice
spacing equal to one,
\begin{eqnarray}
\overline{p}_\mu &=&  \sin p_\mu \; ,\\
 \hat{p}^2 &=& \sum_\mu \left( 2 \sin {p_\mu \over 2} \right)^2 \; , \\
 M_W(p) &=& {r_W\over2} \hat{p}^2 \; .
\end{eqnarray}
The additive mass renormalization $\delta m_B$ 
is obtained by requiring 
$S^{-1}(0,\delta m_B) =0$, i.e.
\be
\Sigma^L(0,\delta m_B,g_0) = \delta m_B.
\ee
This equation can be solved in perturbation theory 
by expanding 
\be
\Sigma^L(0,m_B,g_0) = \sum_{n=1}^\infty g_0^{2n} \,\Sigma^{(n)}.
\ee 
In the following we shall compute 
$\Sigma^{(1)}$ and $\Sigma^{(2)}$ for $r_W=1$, 
gauge group $SU(N)$, and $N_f$ fermionic flavour species. 
We will work in the Feynman gauge.

\section{One-loop result}

In Ref.~\cite{Burgio} we already reported the analytic
one-loop expression for the fermionic self-energy $\Sigma^L$. The
first computation for the Wilson action in Feynman gauge was given
in Ref.~\cite{Gonzalez-Yndurain-Martinelli} and it was subsequently
corrected in Ref.~\cite{Hamber-Wu}.\footnote{Notice, however, that
Eq. (3.15) in Ref.~\cite{Hamber-Wu} contains a misprint: the
correct result is given in Eq. (10b) of
Ref.~\cite{Gonzalez-Yndurain-Martinelli}.}
Our result is expressed in
terms of three purely bosonic constants---their numerical values
are given in Table \ref{costanti_bosoniche}---and
\begin{table}
\begin{center}
\begin{tabular}{cl}
\hline \hline 
$Z_0$ & 0.154933390231060214084837208 \\ $Z_1$ &
0.107781313539874001343391550 \\ $F_0$ & 4.369225233874758 \\
\hline \hline
\end{tabular}
\end{center}
\caption{Numerical values of the three constants $Z_0$, $Z_1$, and
$F_0$. } \label{costanti_bosoniche}
\end{table}
of 12 numerical constants that appear in the presence of Wilson
fermions, whose numerical values are given in Table
\ref{costanti_fermioniche}. Except for $F_0$, 
we report here 20 digits, but many more can be
computed, and indeed {\em must} be computed for 
the numerical implementation of the coordinate-space method, 
see Ref. \cite{LW}. In practice, we have computed all 
constants but $F_0$ with 60-digit precision \cite{Capitani-prep}.
\begin{table}
\begin{center}
\begin{tabular}{cr}
\hline \hline 
${\cal F}(1,0)$ & \hphantom{$-$}  0.08539036359532067914  \\
${\cal F}(1,-1)$ &                0.46936331002699614475  \\
${\cal F}(1,-2)$ & \hphantom{$-$} 3.39456907367713000586  \\
${\cal F}(2,-1)$ &                0.05188019503901136636  \\ 
${\cal F}(2,-2)$ & \hphantom{$-$} 0.23874773756341478520  \\ 
${\cal F}(3,-2)$ &                0.03447644143803223145  \\ 
${\cal F}(3,-3)$ & \hphantom{$-$} 0.13202727122781293085  \\
${\cal F}(3,-4)$ &                0.75167199030295682254  \\ 
$Y_0$ & $-$             0.01849765846791657356  \\
$Y_1$           &       0.00376636333661866811  \\ 
$Y_2$           & \hphantom{$-$} 0.00265395729487879354  \\
$Y_3$           &       0.00022751540615147107  \\ 
\hline \hline
\end{tabular}
\end{center}
\caption{Numerical values of the constants appearing in the
fermionic integrals. } \label{costanti_fermioniche}
\end{table}
These constants  are defined as follows. We define
the integrals
\begin{eqnarray}
\BB(p;n_x,n_y,n_z,n_t) &=& \int_{-\pi}^\pi {d^4k\over (2 \pi)^4}
\, { \hat{k}^{2 n_x}_x \hat{k}^{2 n_y}_y  \hat{k}^{2 n_z}_z
\hat{k}^{2 n_t}_t\over D_B(k,m_b)^{p} } \nonumber \\
\FF(p,q;n_x,n_y,n_z,n_t) &=&
 \int_{-\pi}^\pi {d^4 k\over (2\pi)^4}
 {\hat{k}_x^{2 n_x} \hat{k}_y^{2 n_y} \hat{k}_z^{2 n_z} \hat{k}_t^{2 n_t}\over
     D_F(k,m_f)^{p} D_B(k,m_b)^q},
\end{eqnarray}
where 
\begin{eqnarray}
D_B(k,m_b) & = & \hat{k}^2 + m_b^2, \nonumber \\ 
D_F(k,m_f) & = &
\overline{k}^2 + M_W(k)^2  + m_f^2. \label{regulated}
 \end{eqnarray}
When one of the $n$'s vanishes, we shall not write it.
Then
\begin{eqnarray}
Z_0 &=& \left. {\cal B}(1) \right|_{m_b = 0} \\ Z_1 &=& {1\over 4}
\left. {\cal B}(1;1,1) \right|_{m_b = 0} \\ 
F_0 &=& \lim_{m_b\to 0} [16\pi^2 {\cal B}(2) + \log m^2_b + \gamma_E ], 
\end{eqnarray}
while
\begin{eqnarray}
Y_0 &=&  \lim_{m\to 0} \left[  {\cal F}(2,0) + {1\over 16 \pi^2}
\left( \log m^2 + \gamma_E - F_0\right) \right] \\ Y_1 &=& {1\over
8}\, \left. {\cal F} (1,1;1,1,1) \right|_{m = 0} \nonumber \\ Y_2
&=& {1\over 16}\, \left. {\cal F} (1,1;1,1,1,1)\right|_{m = 0}
\nonumber \\ Y_3 &=& {1\over 16}\, \left. {\cal F}
(1,2;1,1,1)\right|_{m = 0},
\end{eqnarray}
where we have set $m\equiv m_f=m_b$. Also, in Table 
\ref{costanti_fermioniche} the numerical 
values refer to the massless case $m=0$.

\begin{figure}
\begin{center}
\protect\footnotesize
\begin{picture}(54,15)
\put(0,0){\epsfig{figure=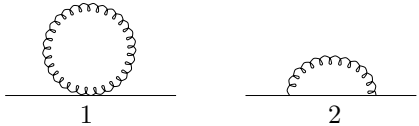,width=0.40\textwidth}}
\put(10,-3.5){1}
\put(43,-3.5){2}
\end{picture} 
\end{center}
\caption{ One-loop diagrams contributing to $\Sigma^L$.
Wavy (solid) lines represent gluons (fermions).
}
\label{1loop}
\end{figure}

At one-loop order 
\be
\Sigma^{(1)} = {N^2-1 \over N}\, \sum_{i=1}^2 c_{i}^{(1)}
\ee     
where $c_{i}^{(1)}$ are the contributions of the two diagrams 
illustrated in Fig.~\ref{1loop}. In terms of the basic 
integrals defined above, they are given by 
\begin{eqnarray*}
c^{(1)}_1 & = &  - Z_0\\
c^{(1)}_2 & = &  {Z_0 \over 2} -  \,\FF(1,0)\; .
\end{eqnarray*}
The numerical values are reported in Table \ref{uno} and compared with 
the results of Ref. \cite{Follana}, obtained by using 
the momentum-space method. The agreement is excellent. 
Summing up the two contributions we obtain
\be
\sum_{i=1}^2 c_{i}^{(1)}  = - \left[ Z_0 + 2 \,\FF(1,0)
\right] \approx - 0.16285705871085078618\; .
\ee 
The constant is in excellent agreement with the result 
of Ref.~\cite{Follana},\par 
\noindent $ \sum_i c_{i}^{(1)} = 0.162857058711(2)$.

\begin{table}
\begin{center}
\begin{tabular}{cl@{}l}
\hline
\hline 
$i$&\multicolumn{1}{c}{$c^{(1)}_{i}$}\\  
\hline
\hline
1 & $-$0.15493339023106  \\ & $-$0.15493339023106021408   \\
2 & $-$0.007923668480(2) \\ & $-$0.00792366847979057210   \\
\hline
\hline
\end{tabular}
\end{center}
\caption{Coefficients $c^{(1)}_{i}$. For each of them
we report in the first line the result of Ref. \cite{Follana},
obtained by means of a momentum-space integration, 
and in the second line our result, obtained by means of 
the coordinate-space method.}
\label{uno}
\end{table}

\section{Two-loop result}

\begin{figure}
\begin{center}
\protect\tiny
\begin{picture}(110,140)
\put(0,0){\epsfig{figure=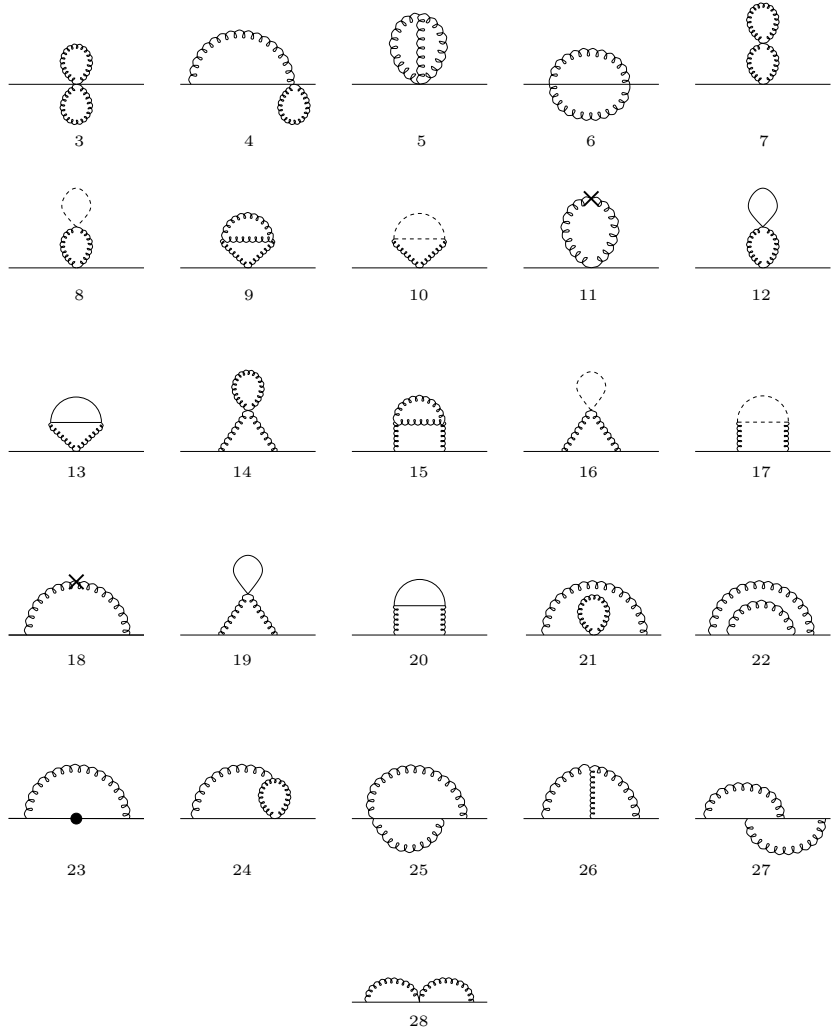,width=0.8\textwidth}}
\put(9,114){3}
\put(31.5,114){4}
\put(54.5,114){5}
\put(77,114){6}
\put(100,114){7}
\put(9,93.5){8}
\put(31.5,93.5){9}
\put(53.5,93.5){10}
\put(76,93.5){11}
\put(99,93.5){12}
\put(8,70){13}
\put(30,70){14}
\put(53.5,70){15}
\put(76,70){16}
\put(99,70){17}
\put(8,45){18}
\put(30,45){19}
\put(53.5,45){20}
\put(76,45){21}
\put(99,45){22}
\put(8,17){23}
\put(30,17){24}
\put(53.5,17){25}
\put(76,17){26}
\put(99,17){27}
\put(53.5,-3){28}
\end{picture} 
\end{center}
\caption{ Two-loop diagrams contributing to $\Sigma^L$.
Wavy (solid, dotted) lines represent gluons (fermions,
ghosts). Crosses denote vertices stemming from the measure part of the
action; the solid circle is the one-loop fermion-mass counterterm.
}
\label{2loop}
\end{figure}

At two loops there are 26 diagrams that have been drawn in
Fig.~\ref{2loop}. They are numbered as 
in Ref.~\cite{Follana} in order to simplify the comparison. 
The $i$-th diagram gives a contribution of the form
\be
D_i \equiv (N^2-1) \left[ c_{1,i}^{(2)} + {1\over N^2} \, c_{2,i}^{(2)} +
{N_f\over N} \, c_{3,i}^{(2)} \right].
 \ee
In Table~\ref{compare} we report the results given in
Ref.~\cite{Follana} and those obtained here by using the configuration-space 
method.
\begin{table}
\begin{center}
\protect\tiny
\begin{tabular}{cr@{}lr@{}lr@{}l}
\hline
\hline 
$i$&\multicolumn{2}{c}{$c^{(2)}_{1,i}$}&\multicolumn{2}{c}{$c^{(2)}_{2,i}$}&
\multicolumn{2}{c}{$c^{(2)}_{3,i}$}\\  
\hline
\hline 
3 & 0&.002000362950707492 & -0&.0030005444260612375& 0&\\ 
& 0&.0020003629507074987148 & $-$0&.0030005444260612480722 & 0& \\ 
\hline 
4 & 0&.00040921361(1) & $-$0&.00061382041(2) & 0& \\ 
& 0&.0004092136068803147865 & $-$0&.0006138204103204721798 & 0&\\ 
\hline 
5 & 0& & 0& & 0& \\ 
& 0& & 0& & 0&\\ 
\hline 
6 & $-$0&.0000488891(8) & 0&.000097778(2) & 0& \\ 
& $-$0&.0000488853119(2)& 0&.000097770623(5)& 0&\\ 
\hline 
7+8+9+10+11 & $-$0&.013927(3) & 0&.014525(2) & 0& \\ 
& $-$0&.01392647740 (2)& 0&.0145250053341618950704 & 0&\\ 
\hline 
12+13 & 0& & 0& & 0&.00079263(8) \\ 
& 0& & 0& & 0&.000792647(2)\\ 
\hline 
14+15+16+17+18 & $-$0&.005753(1) & 0&.0058323(7) & 0& \\ 
& $-$0&.00575248584(1) & 0&.005832127004694453 & 0&\\ 
\hline 
19+20 & 0& & 0& & 0&.000393556(7) \\ 
& 0& & 0& & 0&.000393556(4)\\ 
\hline 
21+22+23 & 0&.000096768(4) & $-$0&.000096768(4) & 0& \\ 
& 0&.0000967648(2) & $-$0&.0000967648(2) & 0&\\ 
\hline 
24 & 0& & 0& & 0& \\ 
& 0& & 0& & 0&\\ 
\hline 
25 & 0&.00007762(1) & $-$0&.00015524(3) & 0& \\ 
& 0&.000077613106(4)& $-$0&.000155226212(8)& 0& \\ 
\hline 
26 & $-$0&.00040000(5) & 0& & 0& \\ 
& $-$0&.00039997586(1)& 0& & 0&\\ 
\hline 
27 & 0& & $-$0&.000006522(1) & 0& \\ 
& 0& & $-$0&.0000065203(1)& 0& \\  
\hline 
28 &  0&.0000078482(5) & $-$0&.000015696(1) & 0& \\ 
& 0&.0000078480652722033294 & $-$0&.0000156961305444066589 & 0& \\ 
\hline 
\hline 
Total & $-$0&.017537(3) & 0&.016567(2) & 0&.00118618(8) \\ 
& $-$0&.0175360218(2) & 0&.0165663304(2) & 0&.001186203(6)\\ 
\hline
\hline
\end{tabular}
\end{center}
\caption{Coefficients $c^{(2)}_{1,i}$, $c^{(2)}_{2,i}$ and
$c^{(2)}_{3,i}$.  For each of them
we report in the first line the result of Ref.~\cite{Follana},
obtained by means of a momentum-space integration, 
and in the second line our result, obtained by means of 
the coordinate-space method.}
\label{compare}
\end{table}
When we have not reported an error, the precision we achieve is much
higher than the reported digits. This occurs in general when the 
diagram is the product of one-loop integrals.  
All results are in agreement with what had been presented
in Ref.~\cite{Follana}. Only for diagram 6 there is apparently
a (very) small underestimation of the error, 
which is negligible in the sum of all contributions.

In Table \ref{compare} diagrams are grouped together
in order to obtain infrared-convergent results. This is
necessary for the numerical implementation of the momentum-space
method. In our case, however, we have followed a different 
strategy considering an infrared regularization. 
This allows us to compute each Feynman diagram separately.
We have used four different infrared regularizations:
\begin{itemize}
\item[(a)] We simply introduce a regulator in the 
propagators as in Eq. \reff{regulated}. Explicitly, for 
the gluon ($\Delta_B(k)$) and for the fermion ($\Delta_F(k)$) propagator
we use:
\begin{eqnarray}
\Delta_B(k) &=& {1\over D_B(k,m)}, \label{gluon-regularized} \\
\Delta_F(k) &=& {- i \bar{k}_\mu \gamma_\mu + M_W(k) \over 
                 D_F(k,m)}\; .
\end{eqnarray}
\item[(b)] We regularize the gluon propagator as in 
Eq. \reff{gluon-regularized}, but use instead the correct Wilson-fermion
propagator
\be
\Delta_F(k) = {- i \bar{k}_\mu \gamma_\mu + M_W(k) + m \over 
                 \bar{k}^2 + (M_W(k) + m)^2}\; .
\ee
\item[(c)] We regularize the Wilson fermion as in (a), but use the 
massless propagator for the gluon.
\item[(d)] We regularize the Wilson fermion as in (b) and the 
massless propagator for the gluon.
\end{itemize}
The result for each diagram will be indicated by $D^{(x)}_i$ 
where $(x)$ refers to the chosen infrared regularization.

In Table~\ref{all} we report the result for each diagram 
computed in the regularization (a).
%
%
%
\def\sfrac#1#2{{\textstyle\frac{#1}{#2}}}
\begin{table}
\begin{center}
\protect\tiny
\begin{tabular}{c l l l l}
\hline
\hline
$i$ & Divergent Part & $c_{1,i}^{(2)}$ & $c_{2,i}^{(2)}$ & $c_{3,i}^{(2)}$\\
\hline
\hline
&&&&\\[-1mm]
7 & $\left(\sfrac{7}{256\pi^2} - \sfrac{3 Z_0}{16\pi^2}\right)\log
m^2$ & --0.01347759288130718 & 0.0145250053341618950704 & 0 \\[1mm] 
8 & $\left(\sfrac{-1}{384\pi^2}\right)\log m^2$ & 0.001000549213311037
& 0 & 0 \\[1mm] 
9 & $\left(\sfrac{-3}{128\pi^2} + \sfrac{7 Z_0}{32\pi^2}\right)\log
m^2$ & --0.00347141693(2) & 0 & 0 \\[1mm] 
10 & $\left(\sfrac{1}{256\pi^2} - \sfrac{Z_0}{32\pi^2}\right)\log m^2$
& 0.000020884775(4) & 0 & 0 \\[1mm] 
11 & $\left(\sfrac{-1}{192\pi^2}\right)\log m^2$ &
0.002001098426622074 & 0 & 0\\[1mm] 
\hline
&&&&\\[-1mm]
12 & $\sfrac{N_f}{N}\left(\sfrac{-1}{32\pi^2} + \sfrac{J[1,
    -1]}{16\pi^2}\right)\log m^2$ &0&0&0.000735684385222637\\[1mm] 
13 & $\sfrac{N_f}{N}\left(\sfrac{1}{32\pi^2} - \sfrac{J[1,
    -1]}{16\pi^2}\right)\log m^2$ & 0 & 0 & 0.000056962(2)\\[1mm] 
\hline
&&&&\\[-1mm]
14 & $\left(\sfrac{7}{512\pi^2} - \sfrac{3 Z_0}{32\pi^2}\right)\log
m^2$  & --0.005494345058160373 & 0.005832127004694453 & 0 \\[1mm] 
15 & $\left(\sfrac{-3}{256\pi^2} + \sfrac{7 Z_0}{64\pi^2}\right)\log
m^2$  & --0.00028296107(1) & 0 & 0 \\[1mm] 
16 & $\left(\sfrac{-1}{768\pi^2}\right)\log m^2$  &
--0.000055474799728634 & 0 & 0 \\[1mm] 
17 & $\left(\sfrac{1}{512\pi^2} - \sfrac{Z_0}{64\pi^2}\right)\log m^2$
& 0.0001912446811(2) & 0 & 0 \\[1mm] 
18 & $\left(\sfrac{-1}{384\pi^2}\right)\log m^2$ &
--0.000110949599457269 & 0 & 0 \\ 
\hline
&&&&\\[-1mm]
19 & $ \sfrac{N_f}{N}\left(\sfrac{1}{32 \pi^2} - \sfrac{J[1, -1]}{16
    \pi^2}\right)$ & 0 &0 &  --0.000040789541774414\\[1mm] 
20 & $ \sfrac{N_f}{N}\left(-\sfrac{1}{32 \pi^2} + \sfrac{J[1, -1]}{16
    \pi^2}\right) $ & 0 & 0 & 0.000434346(4)\\[1mm] 
\hline
&&&&\\[-1mm]
21 & $\left(1-\sfrac{1}{N^2}\right)\sfrac{Z0}{8\pi^2}\log m^2$ &
0.001606284825541242 & --0.001606284825541242 &0\\[1mm] 
22 & $\left(1-\sfrac{1}{N^2}\right)\left(\sfrac{-Z0}{16\pi^2} +
  \sfrac{J[1, 0]}{8\pi^2}\right)\log m^2$ & 0.0005015205(2) &
--0.0005015205(2)&0\\[1mm] 
23 & $\left(1-\sfrac{1}{N^2}\right)\left(\sfrac{-Z0}{16\pi^2} -
  \sfrac{J[1, 0]}{8\pi^2}\right)\log m^2$& --0.002011040454066014 &
0.002011040454066014&0\\[1mm] 
\hline
\hline
\end{tabular}
\end{center}
\caption{Results for the infrared 
regularization (a) at two loops. Graphs whose divergent 
contributions sum up to zero have been grouped together.}
\label{all}
\end{table} 
As expected, the divergences cancel when the appropriate combinations of 
integrals are considered. The finite results are reported in 
Table \ref{compare}.

Then, we consider regularization (b). 
In Table~\ref{check} we report,
for two groups of graphs, the difference $D^{(b)}_i - D^{(a)}_i$,
where the superscripts refer to the chosen regularization. 
Note that, if we use the correct fermion propagator 
there are additional divergences. As expected, the sum of the 
terms reported in Table~\ref{check} vanishes.

As a last check, we have repeated the calculation of diagrams
21, 22, 23 using the regularizations (c) and (d). 
In Table~\ref{P_w} we give the differences $D^{(x)}_i - D^{(a)}_i$
for $x=c,d$. The case $x=c$ is obtained by setting $P_{wil}=0$,
while $x=d$ s obtained by setting $P_{wil}=1$. 
Again, we note that the individual diagrams are different, but the 
sum of the contributions vanishes.

In conclusion, the coordinate-space method is a very efficient tool
for the computation of (infrared-finite and infrared-divergent) 
two-loop Feynman integrals also in the presence of Wilson 
fermions. The method proposed in 
Ref.~\cite{Capitani} really works.

%
%
%
\begin{table} 
\begin{center} 
\protect\footnotesize 
\begin{tabular}{cl} 
\hline 
\hline 
$i$ & 
\\ 
\hline 
\hline 
&\\[-3mm] 
\phantom{space}19\phantom{space} & $ -\sfrac{N_f}{N}\left(
  \sfrac{1}{16 m \pi^2} -  
\sfrac{{\cal F}(1,-1)}{32 \pi^2} - \sfrac{{\cal F}(1,-1)}{8 m \pi^2} -
\sfrac{{\cal F}(1,0)}{4 \pi^2}  
\right) $\\[1mm] 
20 & $\sfrac{N_f}{N}\left( \sfrac{1}{16 m \pi^2} - \sfrac{{\cal
      F}(1,-1)}{32 \pi^2} -   
\sfrac{{\cal F}(1,-1)}{8 m \pi^2} - \sfrac{{\cal F}(1,0)}{4
  \pi^2}\right) $ \\[1mm]  
\hline 
&\\[-3mm] 
21 & $\left(1-\sfrac{1}{N^2}\right)\sfrac{Z_0}{16\pi^2}$\\[1mm] 
22 & $\left(1-\sfrac{1}{N^2}\right)\left(\sfrac{-Z_0}{32\pi^2} + 
\sfrac{{\cal F}(1,0)}{16\pi^2}\right)$\\[1mm] 
23 & $\left(1-\sfrac{1}{N^2}\right)\left(\sfrac{-Z_0}{32\pi^2} - 
\sfrac{{\cal F}(1,0)}{16\pi^2}\right)$\\[1mm] 
\hline 
\hline 
\end{tabular} 
\end{center} 
\caption{Difference $D_i^{(b)} - D_i^{(a)}$.
Graphs whose contributions sum up to zero are grouped together.}
\label{check}
\end{table} 
%
%
%
\begin{table} 
\begin{center} 
\protect\footnotesize 
\begin{tabular}{cl} 
\hline 
\hline 
$i$ &  
\\ 
\hline 
\hline 
&\\[-3mm] 
\phantom{space}21\phantom{space}& $\left(1-\sfrac{1}{N^2}\right)
\left(\sfrac{11 Z_0}{96 \pi^2} + \sfrac{5 P_{wil} Z_0}{96
    \pi^2}\right)$\\[1mm]  
22&$\left(1-\sfrac{1}{N^2}\right)
\left( \sfrac{-11 Z_0}{192 \pi^2} + \sfrac{11 {\cal F}(1,0)}{96
    \pi^2}+P_{wil}\left(-\sfrac{5 Z_0}{192 \pi^2}+\sfrac{5 {\cal
        F}(1,0)}{96 \pi^2}\right)\right)$\\[1mm]  
23 & $\left(1-\sfrac{1}{N^2}\right)
\left( \sfrac{-11 Z_0}{192 \pi^2} - \sfrac{11 {\cal F}(1,0)}{96
    \pi^2}-P_{wil}\left(\sfrac{5 Z_0}{192 \pi^2}+\sfrac{5 {\cal
        F}(1,0)}{96 \pi^2}\right)\right)$\\[1mm] 
\hline 
\hline 
\end{tabular} 
\end{center} 
\caption{Difference $D_i^{(x)} - D_i^{(a)}$. 
  Regularization $(x) = (c)$ is obtained by taking $P_{wil}=0$,
  regularization $(x) = (d)$ by taking $P_{wil}=1$.
  }
\label{P_w}
\end{table} 

\section*{Acknowledgments}

We thank Haris Panagopoulos for an intense correspondence and for
providing us some intermediate unpublished results, which have
been very useful to test our programs.

\end{document}